\documentstyle[epsfig,here,12pt]{elsart}

\newcommand{\be}[1]{\begin{equation} \label{(#1)}}
\newcommand{\ee}{\end{equation}}
\newcommand{\baq}[1]{\begin{eqnarray} \label{(#1)}}
\newcommand{\eaq}{\end{eqnarray}}
\newcommand{\rf}[1]{(\ref{(#1)})}
\newcommand{\ba}{\begin{array}}
\newcommand{\ea}{\end{array}}
\newcommand{\T}{\tilde\tau}
\newcommand{\slashed}[1]{\not\!#1}
\newcommand{\CH}{\tilde\chi}

\def\beq   {\begin{equation}}
\def\eeq   {\end{equation}}
\def\beqd  {\begin{displaymath}}
\def\eeqd  {\end{displaymath}}
\def\beqaa {\begin{eqnarray}}
\def\eeqaa {\end{eqnarray}}

\def\ti  {\tilde}

\def\sz{\ifmmode{\tilde{\chi}^0} \else{$\tilde{\chi}^0$} \fi}
\def\sw{\ifmmode{\tilde{\chi}} \else{$\tilde{\chi}$} \fi}

\def\bold#1{\setbox0=\hbox{$#1$} 
     \kern-.025em\copy0\kern-\wd0 
     \kern.05em\copy0\kern-\wd0 
     \kern-.025em\raise.0433em\box0 }

\begin{document}

\begin{flushright}
   UWThPh-2003-30
 \\WUE-ITP-2003-015
 \\hep-ph/0309340
\end{flushright}

\begin{frontmatter}
  
  \title{A CP asymmetry in 
 $e^+e^-\to \CH^0_i\CH^0_j\to \CH^0_j\tau \T_k$ with tau polarization}

\author{A.~Bartl\thanksref{bartl}},
\author{T.~Kernreiter\thanksref{tkern}},
\author{O.~Kittel\thanksref{kittel}}

\address{(1,2,3) Institut f\"ur Theoretische Physik, Universit\"at Wien, A-1090
Vienna, Austria}
\address{
(3) Institut f\"ur Theoretische Physik und Astrophysik, 
Universit\"at W\"urzburg\\
D-97074 W\"urzburg, Germany}

\thanks[bartl]{ bartl@ap.univie.ac.at}
\thanks[tkern]{tkern@hephy.oeaw.ac.at}
\thanks[kittel]{ kittel@physik.uni-wuerzburg.de}

\bigskip
\noindent

\begin{abstract}
 We propose a CP-odd asymmetry in the supersymmetric process
 $e^+e^-\to \CH^0_i\CH^0_j\to \CH^0_j\tau^{\mp} \T_k^{\pm}$
 by means of the transverse $\tau^{\mp}$ polarization.
 We calculate the asymmetry and cross sections at a future
 linear collider in the 500~GeV c.m.s. energy range with 
 longitudinal polarized beams and high luminosity.
 We work in the Minimal Supersymmetric Standard Model with 
 complex parameters $\mu$, $M_1$ and $A_{\tau}$.
 The asymmetry can reach values up to 60\%. 
 We also estimate the sensitivity for measuring the $\tau$
 polarization necessary to probe the CP asymmetry. 
 \end{abstract}
\end{frontmatter}


\section{Introduction}

In supersymmetric (SUSY) extensions of the Standard Model (SM), 
some parameters can be complex.  
In the neutralino sector of the Minimal Supersymmetric Standard 
Model (MSSM), these are the higgsino mass parameter $\mu$ 
and the gaugino mass parameter $M_1$, while 
$M_2$ can be chosen real by redefining the fields.
In addition, in the scalar tau sector of the MSSM, 
also the trilinear scalar coupling parameter $A_{\tau}$ can 
be complex. The non-zero phases $\varphi_{\mu},\varphi_{M_1}$ and 
$\varphi_{A_{\tau}}$ of these parameters give rise to CP-odd 
observables, which are not present if CP is maintained.
Measurements of such CP-odd observables will allow us to 
determine these phases, in particular also their signs.

In this Letter we consider neutralino production
\be{eq:production}
e^+e^-\to \CH^0_i\CH^0_j; \quad i,j=1,\dots,4
\ee
and the subsequent two-body decay of one neutralino
\be{eq:decay}
\CH^0_i\to \T_m^{\pm} \tau^{\mp}; \quad m=1,2,
\ee
for a fixed $\tau$-polarization. 
We would like to stress that without measuring the 
transverse $\tau^{\mp}$ polarization no sensitivity
to the phase of $A_{\tau}$, $\varphi_{A_{\tau}}$, can
be obtained, because \rf{eq:decay} is a two-body decay.
When summing over the $\tau^-$ polarization, we are sensitive
only to CP violation in the production process~\cite{Bartl:2003ck,olaf}.
The $\tau^-$ polarization is given by \cite{Renard}
\be{eq:polvector}
{\bf P} =\frac{{\rm} {\rm Tr}(\varrho
\ \bold{\sigma})}{{\rm} {\rm Tr}(\varrho)}~,
\ee
with $\varrho$ being the hermitean spin density matrix
of the $\tau^-$
and $\sigma_i$ the Pauli matrices.
We use a convention for ${\bf P}=(P_1,P_2,P_3)$ where
the component $P_3$ is the longitudinal polarization and 
$P_1$ is the transverse polarization in the plane formed by 
${\bf p}_{e^-}$ and ${\bf p}_{\tau}$.
The component $P_2$ is the polarization perpendicular to
${\bf p}_{\tau}$ and ${\bf p}_{e^-}$ and
is proportional to the triple-product
\be{eq:triplepol}
{\bf s}_{\tau}\cdot({\bf p}_{\tau}\times {\bf p}_{e^-})~,
\ee
where ${\bf s}_{\tau}$ is the $\tau^-$ spin 3-vector.
Since under time reversal the triple-product changes sign,
the transverse $\tau^-$ polarization $P_2$ is a T-odd observable.
Due to CPT invariance, $P_2$ is actually a CP-odd observable if 
absorbtive phases from  final-state interactions are 
neglected. 

In this Letter we study the asymmetry 
\be{eq:asy}
{\mathcal A}_{\rm CP}=\frac{1}{2}(P_2-\bar{P}_2)~,
\ee
which is CP-odd, even if absorbtive phases can not be neglected.
In Eq.~\rf{eq:asy}, $\bf P$ denotes the $\tau^-$
polarization vector in the decay $\CH^0_i\to \T_m^+ \tau^-$
and $\bar{\bf P}$ denotes the 
$\tau^+$ polarization vector in the decay $\CH^0_i\to \T_m^- \tau^+$.
In Born approximation it follows from Eq.~\rf{eq:asy} 
that ${\mathcal A}_{\rm CP} = P_2$.

In Section \ref{Definitions} we briefly review
stau mixing in the MSSM and define the part of the interaction
Lagrangian which is relevant for our analysis.
In Section \ref{matrixele} we define the 
$\tau$ spin density matrix $\varrho$ and give 
the anlytical formulae. In Section \ref{taupolarization} we discuss the 
qualitative properties of the asymmetry ${\mathcal A}_{\rm CP}$. 
We present numerical results for $ e^+e^-\to \CH^0_1\T_1\tau$ 
in Section \ref{numerics}. 
We  summarize and conclude in Section \ref{conclusion}.

\section{Stau mixing and Lagrangian\label{Definitions}}

We give a short account of $\T_L - \T_R$ mixing for
complex $\mu =|\mu|e^{i\varphi_{\mu}}$,  
$A_{\tau}=|A_{\tau}|e^{i\varphi_{A_{\tau}}}$ and
$M_1 =|M_1|e^{i\varphi_{M_1}}$. 
The masses and couplings of the $\tau$-sleptons follow from the 
hermitian $2 \times 2$ mass matrix which in the basis 
$(\ti \tau_L, \ti \tau_R)$ reads \cite{elru,guha}
\begin{equation}
{\mathcal{L}}_M^{\T}= -(\T_L^{\dagger},\, \T_R^{\dagger})
\left(\begin{array}{ccc}
M_{\T_{LL}}^2 & e^{-i\varphi_{\T}}|M_{\T_{LR}}^2|\\[5mm]
e^{i\varphi_{\T}}|M_{\T_{LR}}^2| & M_{\T_{RR}}^2
\end{array}\right)\left(
\begin{array}{ccc}
\T_L\\[5mm]
\T_R \end{array}\right),
\label{eq:mm}
\end{equation}
with
\begin{eqnarray} 
M_{\T_{LL}}^2 & = & M_{\tilde L}^2+(-\frac{1}{2}+\sin^2\Theta_W)
\cos2\beta \ m_Z^2+m_{\tau}^2 ,\label{eq:mll}\\[3mm]
M_{\T_{RR}}^2 & = & M_{\tilde E}^2-\sin^2\Theta_W\cos2\beta \
m_Z^2+m_{\tau}^2 ,\label{eq:mrr}\\[3mm]
 M_{\T_{RL}}^2 & = & (M_{\T_{LR}}^2)^{\ast}=
  m_{\tau}(A_{\tau}-\mu^{\ast}  
 \tan\beta), \label{eq:mlr}
\end{eqnarray}
\begin{equation}
\varphi_{\T}  = \arg\lbrack A_{\tau}-\mu^{\ast}\tan\beta\rbrack ,
\label{eq:phtau}
\end{equation}
where 
$m_{\tau}$ is the mass of the $\tau$-lepton, 
$\Theta_W$ is the weak mixing angle,
$\tan\beta=v_2/v_1$ with $v_1 (v_2)$ being the vacuum 
expectation value of the Higgs field $H_1^0 (H_2^0)$,
and $M_{\ti L}$, $M_{\ti E}, A_{\tau}$ are the soft
SUSY--breaking parameters of the $\T_i$ system. 
The $\T$ mass eigenstates are 
$(\tilde\tau_1, \tilde \tau_2)=(\T_L, \T_R)
{{\mathcal R}^{\T}}^T$ with 
 \begin{equation}
{\mathcal R}^{\T}=\left( \begin{array}{ccc}
e^{i\varphi_{\T}}\cos\theta_{\T} & 
\sin\theta_{\T}\\[5mm]
-\sin\theta_{\T} & 
e^{-i\varphi_{\T}}\cos\theta_{\T}
\end{array}\right),
\label{eq:rtau}
\end{equation}
and
\begin{equation}
\cos\theta_{\T}=\frac{-|M_{\T_{LR}}^2|}{\sqrt{|M_{\T _{LR}}^2|^2+
(m_{\T_1}^2-M_{\T_{LL}}^2)^2}},\quad
\sin\theta_{\T}=\frac{M_{\T_{LL}}^2-m_{\T_1}^2}
{\sqrt{|M_{\T_{LR}}^2|^2+(m_{\T_1}^2-M_{\T_{LL}}^2)^2}}.
\label{eq:thtau}
\end{equation}
The mass eigenvalues are
\begin{equation}
 m_{\T_{1,2}}^2 = \frac{1}{2}\left((M_{\T_{LL}}^2+M_{\T_{RR}}^2)\mp 
\sqrt{(M_{\T_{LL}}^2 - M_{\T_{RR}}^2)^2 +4|M_{\T_{LR}}^2|^2}\right).
\label{eq:m12}
\end{equation}

The part of the interaction Lagrangian of the MSSM
relevant to study the decay \rf{eq:decay} reads 
(in our notation and conventions
we follow closely \cite{Haber:1984rc,thomas}):
\be{eq:LagStauchi}
{\mathcal L}_{\ti \tau \tau\CH^0}=  \ti \tau_k \bar \tau
(b^{\ti \tau}_{ki} P_L+a^{\ti \tau}_{ki} P_R)\CH^0_i + {\rm h.c.}~,
\; i =1,\dots,4~, \; k=1,2~,
\ee
with
\be{eq:coupl1}
a_{ki}^{\ti \tau}=
g({\mathcal R}^{\ti \tau}_{kn})^{\ast}{\mathcal A}^{\tau}_{in},\qquad 
b_{ki}^{\ti f}=
g({\mathcal R}^{\ti \tau}_{kn})^{\ast}
{\mathcal B}^{\tau}_{in},\qquad
(n=L,R)
\ee
\begin{equation}
{\mathcal A}^{\tau}_i=\left(\begin{array}{ccc}
f^{\tau}_{Li}\\[2mm]
h^{\tau}_{Ri} \end{array}\right),\qquad 
{\mathcal B}^{\tau}_i=\left(\begin{array}{ccc}
h^{\tau}_{Li}\\[2mm]
f^{\tau}_{Ri} \end{array}\right),
\label{eq:coupl2}
\end{equation}
and
\baq{eq:coupl}
h^\tau_{Li}&=& (h^\tau_{Ri})^{\ast}=Y_{\tau} N_{i3}^{\ast},\nonumber\\
f^\tau_{Li}&=& -\frac{1}{\sqrt{2}}(\tan\Theta_W 
N_{i1}+N_{i2}),\nonumber\\
f^\tau_{Ri}&=& \sqrt{2}\tan\Theta_W N_{i1}^{\ast},
\eaq
where $Y_{\tau}=m_{\tau}/\sqrt{2}m_W \cos\beta$, 
$P_{L,R}=1/2(1\mp \gamma_5)$ and $g$ 
is the weak coupling constant. 
$N$ is the 4$\times$4 unitary neutralino mixing matrix, 
which diagonalizes
the neutral gaugino-higgsino mass matrix $Y_{\alpha\beta}$, 
$N_{i \alpha}^*Y_{\alpha\beta}N_{k\beta}^{\ast}=
m_{\tilde{\chi}^0_i}\delta_{ik}$,
in the basis ($\tilde{B},
\tilde{W}^3, \tilde{H}^0_1, \tilde{H}^0_2$) \cite{Haber:1984rc}.
The part of the Lagrangian for the neutralino production  
\rf{eq:production} can be found e.g. in \cite{gudi,Bartl:hp}.

\section{Tau spin density matrix  \label{matrixele}}

The unnormalized, hermitean, $2\times 2$ spin density matrix
of the $\tau^-$ is defined by:
\be{eq:spindesity}
\varrho^{\lambda_k\lambda'_k}\equiv
\int (|{\mathcal M}|^2)^{\lambda_k\lambda'_k}
{\rm dLips}~,
\ee
where $\mathcal M$ and ${\rm dLips}$ are the
amplitude squared and the Lorentz invariant phase space element, 
respectively, for the whole process of neutralino production 
\rf{eq:production} and decay \rf{eq:decay}. 
The $\tau^-$ helicities are denoted by $\lambda_k$ and $\lambda'_k$.
In the spin density matrix formalism (as used e.g. in \cite{spin,gudi})
the amplitude squared  can be written as
\be{eq:matrixelement}
(|{\mathcal M}|^2)^{\lambda_k\lambda'_k}=2
|\Delta(\CH^0_i)|^2~
\sum_{\lambda_i,\lambda'_i}~
{(\rho_P)}^{\lambda_i\lambda'_i}~
{(\rho_{D})}_{\lambda'_i\lambda_i}^{\lambda_k\lambda'_k}.
\ee
It is composed of the unnormalized spin density matrices 
$\rho_{P}$ for the production
\rf{eq:production} and $\rho_{D}$ for the decay \rf{eq:decay},
the propagator  
$ \Delta(\ti\chi^0_i ) = 
1/[p^2_{\chi_i} -m^2_{\chi_i}
+im_{\chi_i}\Gamma_{\chi_i}]$, with 
$p_{\chi_i}, m_{\chi_i}, \Gamma_{\chi_i}$
being the four-momenta, masses and widths of the decaying neutralino, 
respectively.
$\rho_{P}$ and $\rho_{D}$ carry the
helicity indices $\lambda_i,\lambda'_i$ of the neutralinos 
and/or the helicity indices $\lambda_k,\lambda'_{k}$ of the $\tau^-$.
The factor 2 in Eq.~\rf{eq:matrixelement} is due to the
summation of the $\CH^0_j$ helicities, whose decay is not considered.
We introduce a set of spin basis vectors
$s^a_{\chi_i}\;(a=1,2,3)$ 
for the neutralino $ \ti\chi^0_i$, 
which fulfill the orthonormality relations 
$s^a_{\chi_i}\cdot s^b_{\chi_i}=-\delta^{ab}$ and
$s^a_{\chi_i}\cdot p_{\chi_i}=0$.
Then the spin density matrices can be 
expanded in terms of the Pauli matrices:
\baq{eq:rho}
{(\rho_P)}^{\lambda_i\lambda'_i}&=&P~\delta_{\lambda_i\lambda'_i}+
\Sigma^a_P~\sigma^a_{\lambda_i\lambda'_i}~,\nonumber \\
{(\rho_{D})}_{\lambda'_i\lambda_i}^{\lambda_k\lambda'_k}&=&
\left[D^{\lambda_k\lambda'_k}~\delta_{\lambda'_i\lambda_i}+
(\Sigma^a_{D})^{\lambda_k\lambda'_k}~\sigma^a_{\lambda'_i\lambda_i}~
\right]~.
\eaq
The analytical formulae of $P$ and $\Sigma^a_P$ can be found in 
\cite{gudi}.
Introducing also a set of spin basis vectors $s^b_{\tau}$
for the $\tau^-$,  $D^{\lambda_k\lambda'_k}$
and $(\Sigma_{D}^a)^{\lambda_k\lambda'_k}$ can be expanded:
\baq{eq:D1}
D^{\lambda_k\lambda'_k} &=& D \delta_{\lambda_k\lambda'_k}+
D^b \sigma^b_{\lambda_k\lambda'_k}~,\label{eqD1}\\
%
(\Sigma_{D}^a)^{\lambda_k\lambda'_k} 
&=& \Sigma_{D}^a \delta_{\lambda_k\lambda'_k}+
\Sigma_{D}^{ab} \sigma^b_{\lambda_k\lambda'_k}~. \label{eq:Sigma1}
\eaq
The expansion coefficient are given by
\be{eq:d1}
D={\rm Re}({b^{\ti \tau}_{mi}}^*a^{\ti \tau}_{mi})
m_{\tau} m_{\CH_i} 
+\frac{1}{2}(|b^{\ti \tau}_{mi}|^2+|a^{\ti \tau}_{mi}|^2) 
(p_{\tau}\cdot p_{\CH_i})~,
\ee
\be{eq:db1}
D^b=\frac{1}{2}m_{\tau}(|b^{\ti \tau}_{mi}|^2-
|a^{\ti \tau}_{mi}|^2)(p_{\CH_i}\cdot s^b_{\tau})~,
\ee
\be{eq:sa1}
\Sigma_{D}^a=\frac{1}{2}m_{\CH_i}(|a^{\ti \tau}_{mi}|^2-
|b^{\ti \tau}_{mi}|^2)(p_{\tau}\cdot s^a_{\CH_i})~,
\ee
\baq{eq:sab1}
\Sigma_{D}^{ab}&=&{\rm Re}({b^{\ti \tau}_{mi}}^*a^{\ti \tau}_{mi})
(p_{\tau}\cdot s^a_{\CH_i})
(p_{\CH_i}\cdot s^b_{\tau})
{}\nonumber\\[3mm]
{}&&-(s^a_{\CH_i}\cdot s^b_{\tau})
\lbrack\frac{1}{2}(|b^{\ti \tau}_{mi}|^2+
|a^{\ti \tau}_{mi}|^2)m_{\tau} m_{\CH_i}+
{\rm Re}({b^{\ti \tau}_{mi}}^*a^{\ti \tau}_{mi})
(p_{\tau}\cdot p_{\CH_i})\rbrack
{}\nonumber\\[3mm]
{}&&+{\rm Im}({b^{\ti \tau}_{mi}}^*a^{\ti \tau}_{mi})
\epsilon^{\mu\nu\rho\sigma}
{p_{\tau}}_{\mu} {p_{\CH_i}}_{\nu}{s^a_{\CH_i}}_{\rho}
{s^b_{\tau}}_{\sigma}~,
\eaq
with $\epsilon^{0123}=1$.
The last term in Eq.~\rf{eq:sab1} contains the triple 
product \rf{eq:triplepol}. This term is proportional
to ${\rm Im}({b^{\ti \tau}_{mi}}^*a^{\ti \tau}_{mi})$  
and is therefore sensitive to the phases $\varphi_{A_{\tau}},
\varphi_{\mu}$ and $\varphi_{M_1}$. 
Inserting the density matrices 
of Eq.~\rf{eq:rho} into Eq.~\rf{eq:matrixelement}
yields
\be{eq:matrixelement2}
(|{\mathcal M}|^2)^{\lambda_k\lambda'_k}=
4 |\Delta(\CH^0_i)|^2~
\lbrack (P D + \Sigma^a_P \Sigma^a_{D})\delta_{\lambda_k\lambda'_k}+
(P D^b+\Sigma^a_P\Sigma^{ab}_{D})
\sigma^b_{\lambda_k\lambda'_k})\rbrack~.
\ee

\section{Transverse tau polarization and CP asymmetry \label{taupolarization}}

From Eqs.~\rf{eq:polvector} and \rf{eq:matrixelement2} 
we obtain for the transverse $\tau^-$ polarization
\be{eq:polasy1}
P_2= \frac{\int~|\Delta(\CH^0_i)|^2~\Sigma^a_P \Sigma^{a2}_{D}
~{\rm dLips}}
{\int ~|\Delta(\CH^0_i)|^2 ~P D~{\rm dLips}}~,
\ee
which follows because in the numerator we have used 
$\int~|\Delta(\CH^0_i)|^2
~P D^2~{\rm dLips}=0$ and in the denominator we have used
$\int~|\Delta(\CH^0_i)|^2~\Sigma^a_P \Sigma^a_D
~{\rm dLips}=0$.
As can be seen from Eq.~\rf{eq:polasy1}, 
$P_2$ is proportional to the spin correlation term
$\Sigma^{a2}_{D}$, Eq.~\rf{eq:sab1}, which contains the 
CP-sensitive part ${\rm Im}({b^{\ti \tau}_{mi}}^*a^{\ti \tau}_{mi})
\epsilon^{\mu\nu\rho\sigma}
{p_{\tau}}_{\mu} {p_{\CH_i}}_{\nu}{s^a_{\CH_i}}_{\rho}
{s^2_{\tau}}_{\sigma}$.
In order to study the dependence of  $P_2$  on the parameters, 
we expand
\baq{eq:Im}
&&{\rm Im}({b^{\ti \tau}_{1i}}^*a^{\ti \tau}_{1i})= g^2\lbrack
\cos^2\theta_{\T} Y_{\tau}{\rm Im}(f^{\tau}_{Li} N_{i3})+
\sin^2\theta_{\T} Y_{\tau}\sqrt{2}\tan\Theta_W{\rm Im}(N_{i1}N_{i3})
\nonumber\\[3mm]
&&{}+\sin^2\theta_{\T}\cos^2\theta_{\T}(Y^2_{\tau}{\rm Im}
(N_{i3}N_{i3}e^{i\varphi_{\T}})+\sqrt{2}\tan\Theta_W
{\rm Im}(f^{\tau}_{Li} N_{i1}e^{-i\varphi_{\T}}))\rbrack~,
\eaq
using Eqs.~\rf{eq:coupl1}-\rf{eq:coupl} for $m=1$.
If CP violation is solely due to $\varphi_{A_{\tau}}\neq 0$ (mod $\pi$),
we find from \rf{eq:Im} that $P_2\propto \sin 2\theta_{\T} \sin\varphi_{\T}$.
We note that the dependence of $\varphi_{\T}$
on $\varphi_{A_{\tau}}$ is weak if 
$|A_{\tau}|\ll |\mu|\tan\beta$, see Eq.~(\ref{eq:phtau}). Thus, we expect
that $P_2$ increases with increasing $|A_{\tau}|$.

Details concerning phase space and  kinematics necessary for the
calculation of  
$P_2$ from Eq.~\rf{eq:polasy1} can be found in \cite{olaf}.
The $\tau^-$ spin vectors $s^b_{\tau}$ are chosen by:
\be{eq:polvec}
s^1_{\tau}=\left(0,\frac{{\bf s}_2\times{\bf s}_3}
{|{\bf s}_2\times{\bf s}_3|}\right),\quad
s^2_{\tau}=\left(0,
\frac{{\bf p}_{\tau}\times{\bf p}_{e^-}}
{|{\bf p}_{\tau}\times{\bf p}_{e^-}|}\right),\quad
s^3_{\tau}=\frac{1}{m_{\tau}}
\left(|{\bf p}_{\tau}|, \frac{E_{\tau}}{|{\bf p}_{\tau}|}{\bf p}_{\tau} \right)~.
\ee

In order to measure $P_2$ and the CP asymmetry 
${\mathcal A}_{\rm CP}$, Eq.~\rf{eq:asy},
the $\tau^-$ from the decay \rf{eq:decay}
and the $\tau^+$ from the subsequent $\T_m^+$
decay, $\T_m^+\to\CH_1^0\tau^+$, have to be distinguished.
This can be accomplished by measuring the energies
of the $\tau$'s and making use of their different energy distributions
\cite{olaf}.

A potentially large background may be due to stau production
$ e^+e^-\to\tilde\tau_l^+ \tilde\tau_m^-
\to\tau^+\tau^-\tilde\chi^0_1\tilde\chi^0_1$.
However, these reactions would generally lead to ''two-sided
events``, whereas the events from
$ e^+e^-\to\tilde\chi^0_1 \tilde\chi^0_i
\to\tau^+\tau^-\tilde\chi^0_1\tilde\chi^0_1$
are ''one-sided events``. Moreover, the background
reaction is CP-even and will not give rise to a CP asymmetry, because
the staus are scalars with a two-body decay.

\section{Numerical results\label{numerics}}

We present numerical results for
$e^+e^-\to \CH^0_1\CH^0_2$ and the subsequent decay of the neutralino
into the lightest stau $\CH^0_2\to \T_1\tau $
for a linear collider (LC) with
$\sqrt{s}=500$ GeV and longitudinal polarized beams with
$(P_{e^-},P_{e^+})=(0.8,-0.6)$ or $(P_{e^-},P_{e^+})=(-0.8,0.6)$.
This choice favors right or left selectron exchange in the neutralino
production process, respectively \cite{gudi}.
We study the dependence of the asymmetry ${\mathcal A}_{\rm CP}$ 
and the production cross sections 
$\sigma\equiv \sigma_p(e^+e^-\to \CH^0_1\CH^0_2)\times 
BR(\CH^0_2\to\T_1^+\tau^-)$ on the parameters 
$\varphi_{\mu}$, $|\mu|$, $\varphi_{M_1}$, $|M_1|$,
$\varphi_{A_{\tau}}$, $|A_{\tau}|$ and $\tan\beta$.

For the calculations we assume 
$|M_1|=5/3 \tan^2\Theta_W M_2$, 
$m_{\tau}=0$ and use in Eqs.~(\ref{eq:mll})  and~(\ref{eq:mrr}) 
the renormalization group equations (RGEs)
for the selectron masses \cite{hall}, $M_{\ti L}^2 = m^2_0+0.79 M^2_2$ 
and $M_{\ti E}^2 = m^2_0+0.23 M^2_2$, taking $m_0=100$~GeV. 
The size of $|A_{\tau}|$ is restricted due to the tree-level
vacuum stability conditions \cite{casas}. The restrictions on the masses of the
SUSY particles are $m_{\CH^{\pm}_1}>104$~GeV, $m_{\T_1}>100$~GeV and 
$m_{\T_1}>m_{\CH^0_1}$. For the calculation of  BR$(\CH^0_2\to\T_1^+\tau^-)$
we concentrate on the parameter domain where two-body decays are allowed
and neglect three-body decays. We consider the two-body decays
\begin{eqnarray}
\tilde\chi^0_2 &\to& \tilde\tau_{m}\tau,~
\tilde\ell_{R,L}\ell,~ 
\tilde\chi^0_2 Z,~
\tilde\chi^{\mp}_n W^{\pm},~
\tilde\chi^0_1 H_1^0,
\quad \ell=e,\mu, \quad m,n=1,2,
\end{eqnarray}
with $H_1^0$ being the lightest neutral Higgs boson.
The Higgs mass parameter $m_{A}$ is chosen as $m_{A}=1000$~GeV,
which means that explicit CP violation is not important
for the lightest Higgs state \cite{ref3}.
Furthermore,
the neutralino decays into charginos and the charged Higgs bosons 
$\tilde\chi^0_2 \slashed{\to} \tilde\chi^{\pm}_n H^{\mp}$,
as well as decays into the heavy neutral Higgs
bosons $\tilde\chi^0_2 \slashed{\to} \tilde\chi^0_1~H_{2,3}^0$,
are ruled out in this scenario.

\begin{figure}
			\begin{minipage}{0.47\textwidth}
 \begin{picture}(100,230)(0,0)
	\put(0,0){\includegraphics{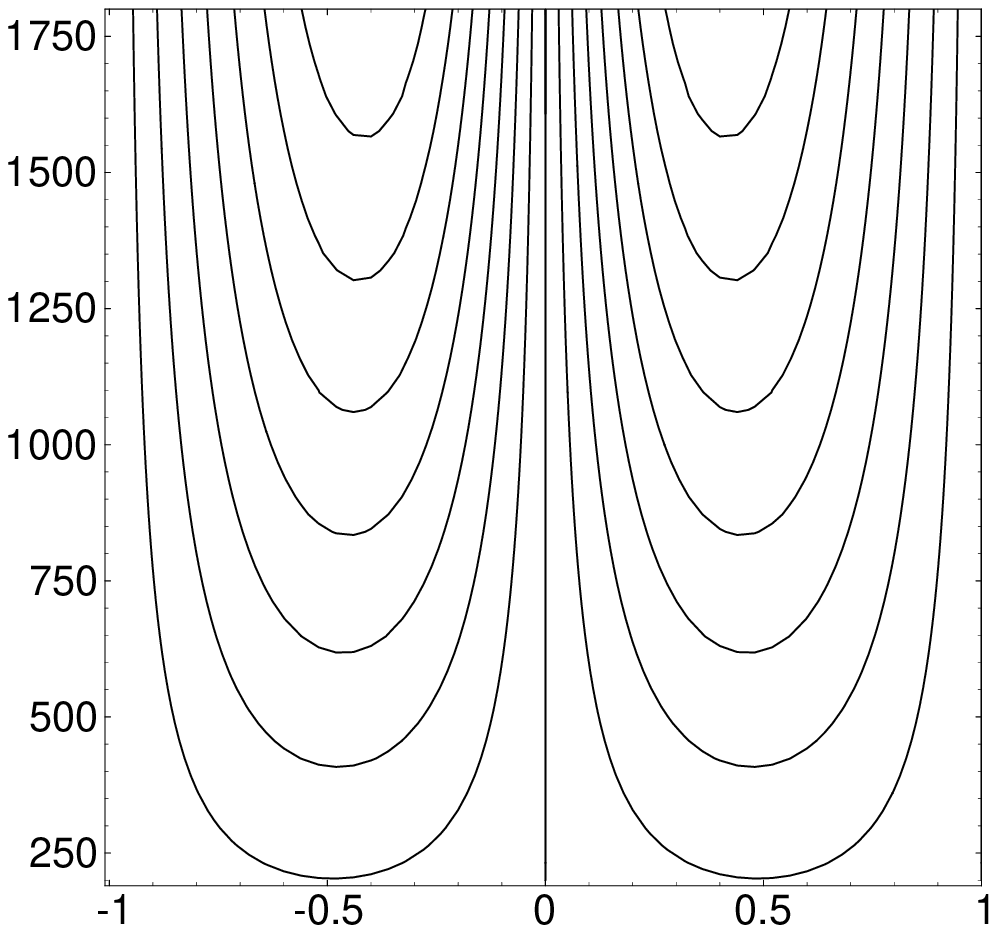}}
	\put(80,210){\fbox{${\mathcal A}_{\rm CP}$ in \% }}
	\put(170,-5){$\varphi_{A_{\tau}}/\pi$}
	\put(0,210){$ A_{\tau}/$ GeV}
	\put(60,23){\footnotesize{$-2$}}
	\put(60,50){\footnotesize{$-4$}}
	\put(61,75){\footnotesize{$-6$}}
	\put(61,100){\footnotesize{$-8$}}
	\put(60,125){\footnotesize{$-10$}}
	\put(60,150){\footnotesize{$-12$}}
	\put(61,185){\footnotesize{$-14$}}
	\put(150,23){\footnotesize{$2$}}
	\put(150,50){\footnotesize{$4$}}
	\put(150,75){\footnotesize{$6$}}
	\put(150,100){\footnotesize{$8$}}
	\put(146,125){\footnotesize{$10$}}
	\put(145,150){\footnotesize{$12$}}
	\put(144,185){\footnotesize{$14$}}
	\put(102,22){\footnotesize{$0$}}
 \end{picture}
\vspace*{.25cm}
\caption{Contour lines of  ${\mathcal A}_{\rm CP}$ in Eq.~\rf{eq:asy}
for $M_2=200$~GeV, $|\mu|=250$~GeV, 
$\tan\beta=5$, $\varphi_{M_1}=\varphi_{\mu}=0$ and 
$(P_{e^-},P_{e^+})=(0.8,-0.6)$.
\label{fig1}}
\end{minipage}
\hspace*{0.5cm}
\begin{minipage}{0.47\textwidth}
 \begin{picture}(100,230)(0,0)
	\put(0,0){\includegraphics{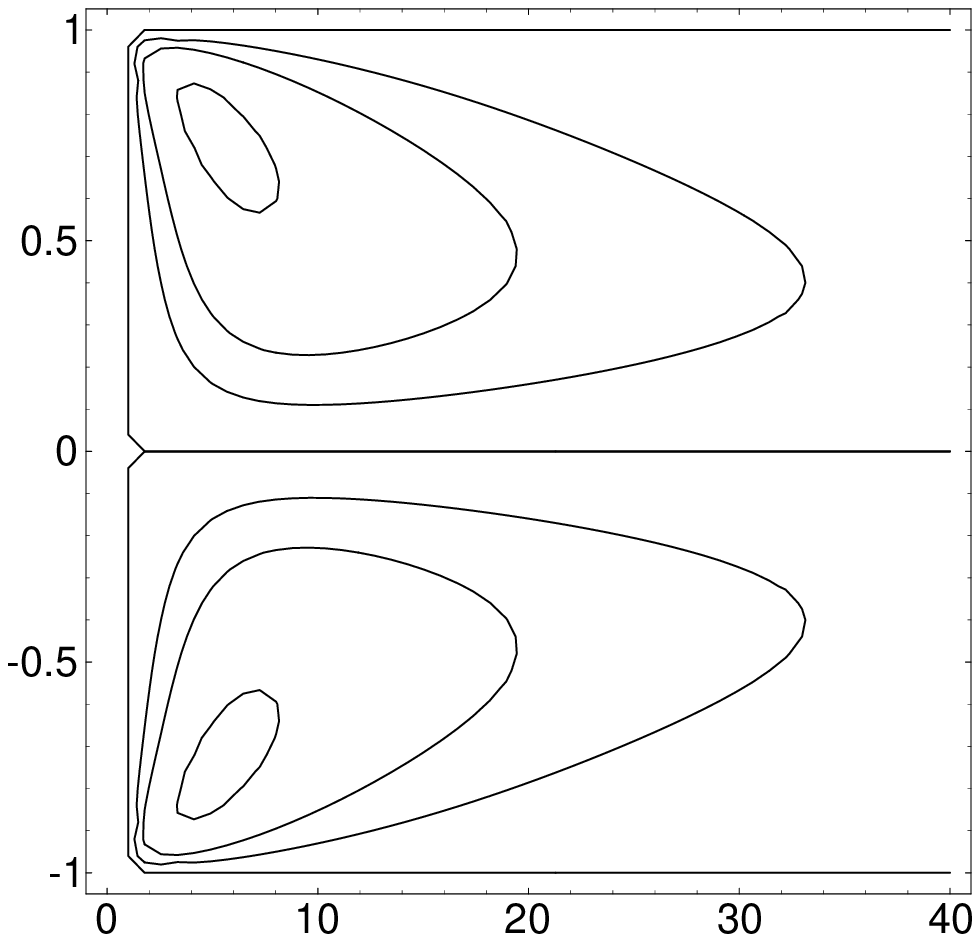}}
	\put(80,210){\fbox{${\mathcal A}_{\rm CP}$ in \% }}
	\put(170,-5){$\tan\beta$}
	\put(0,210){$ \varphi_{M_{1}}/\pi$ }
	\put(35,40){\footnotesize{$-20$}}
	\put(85,62){\footnotesize{$-10$}}
	\put(144,70){\footnotesize{$-5$}}
	\put(175,110){\footnotesize{$0$}}
		\put(45,165){\footnotesize{$20$}}
		\put(90,145){\footnotesize{$10$}}
		\put(154,140){\footnotesize{$5$}}
 \end{picture}
\vspace*{.25cm}
\caption{Contour lines of  ${\mathcal A}_{\rm CP}$ in Eq.~\rf{eq:asy}
for $A_{\tau}=1$~TeV, $M_2=300$~GeV, $|\mu|=250$~GeV, 
$\varphi_{A_{\tau}}=\varphi_{\mu}=0$
and $(P_{e^-},P_{e^+})=(0.8,-0.6)$.
\label{fig2}}
\end{minipage}
\vspace*{.5cm}
\end{figure}

In Fig.~\ref{fig1} we show the contour lines for ${\mathcal A}_{\rm CP}$
in the $\varphi_{A_{\tau}}$-$|A_{\tau}|$ plane.
As can be seen ${\mathcal A}_{\rm CP}$ is proportional to
$\sin 2\theta_{\T} \sin\varphi_{\T}$, which is expected from Eq.~\rf{eq:Im}.  
${\mathcal A}_{\rm CP}$ increases with increasing
$|A_{\tau}|\gg |\mu|\tan\beta$, which also follows from  Eq.~\rf{eq:Im}. 
Furthermore, in the parameter region shown the cross section 
$\sigma$ varies between 20~fb and 30~fb.

In Fig.~\ref{fig2} we show the dependence
of ${\mathcal A}_{\rm CP}$ on $\tan\beta$ and $\varphi_{M_1}$.
Large values up to $\pm 20\%$ are obtained for $\tan\beta\approx 5$. 
Note that these values are obtained for $\varphi_{M_1}\approx \pm0.8\pi$
rather than for maximal 
$\varphi_{M_1}\approx\pm0.5\pi$.
This is due to the 
complex interplay of the spin correlation terms
in Eq.~\rf{eq:polasy1}. In the region shown in Fig.~\ref{fig2},
the cross section $\sigma$ varies between 10~fb and 30~fb.

\begin{figure}
\begin{picture}(120,220)(0,0)
\put(0,0){\includegraphics{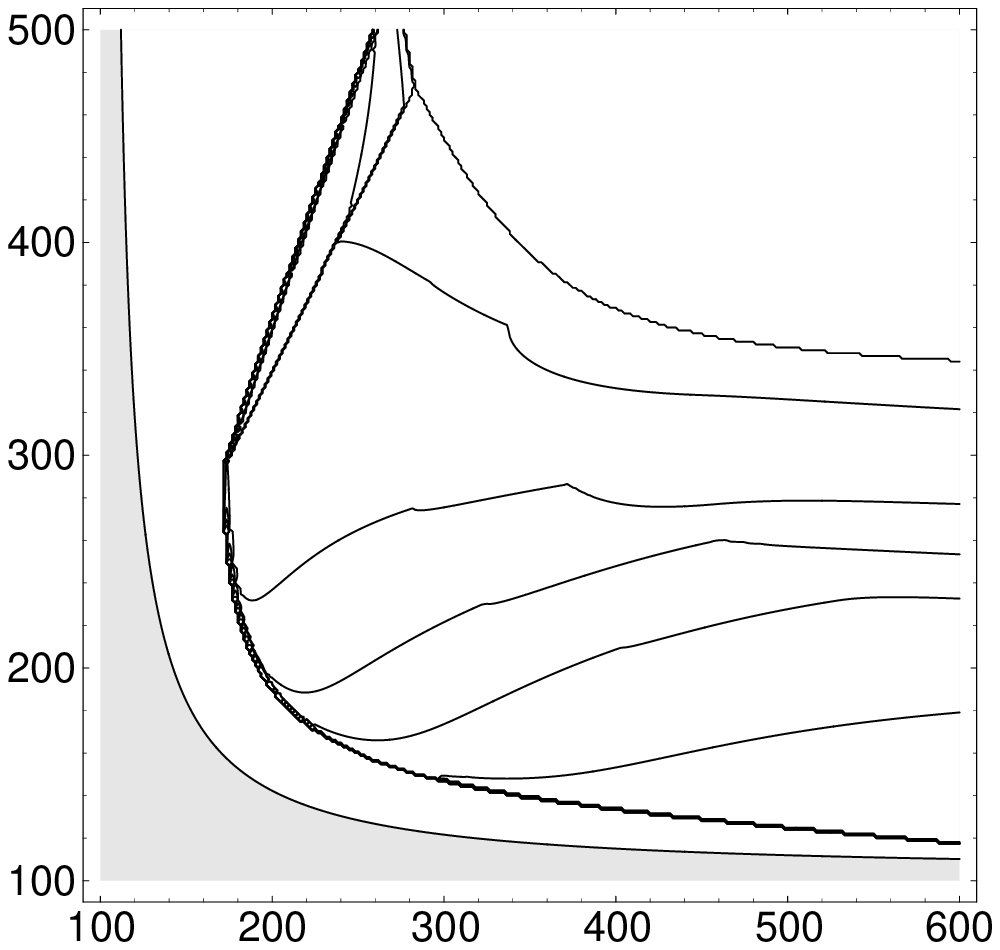}}
   \put(165,-8){$|\mu|$/GeV}
	\put(-5,210){$M_2$/GeV}
	\put(42,210){\fbox{$\sigma(e^+e^-\to \CH^0_1\T_1^+\tau^-)$ in fb }}
	\put(170,55){\footnotesize{100}}
	\put(135,72){\footnotesize{50}}
	\put(110,82){\footnotesize{25}}
	\put(85,98){\footnotesize{10}}
	\put(80,146){\footnotesize{1}}
		\put(20,-8){Fig.~\ref{fig3}a}
\put(225,0){\includegraphics{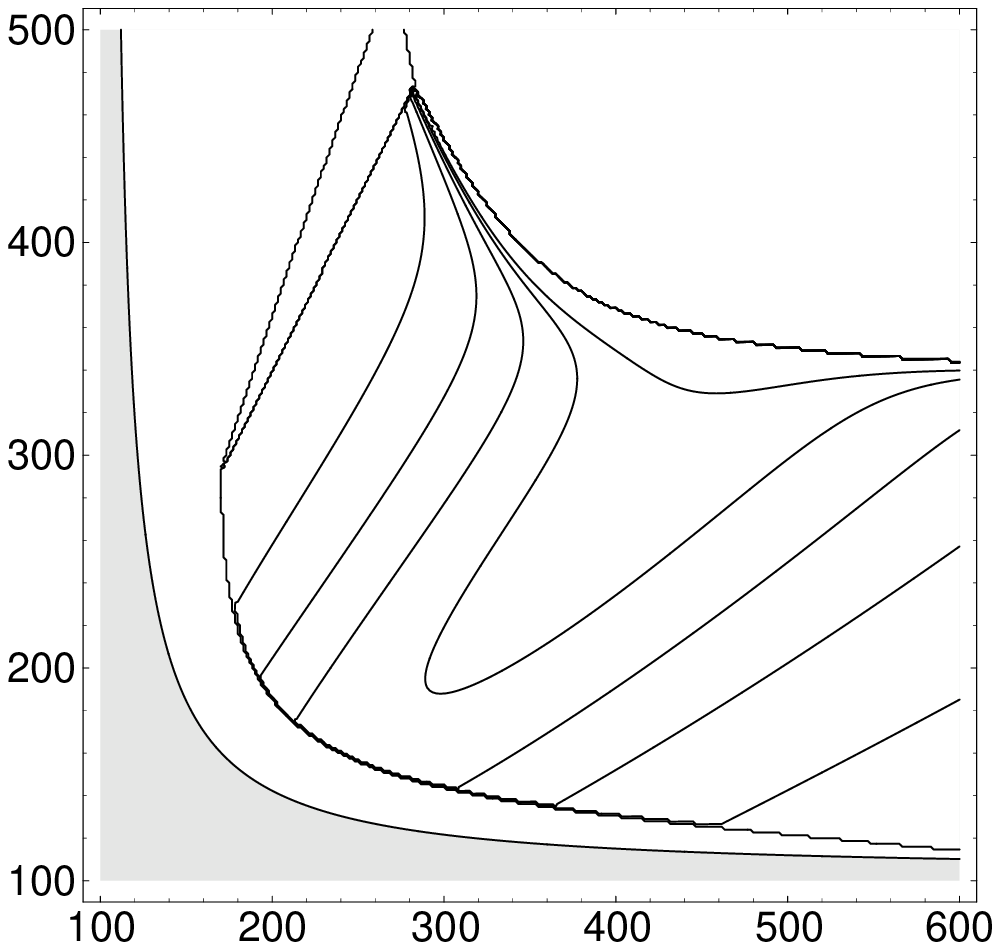}}
   \put(390,-8){$|\mu|$/GeV}
	\put(225,210){$M_2 /$GeV}
	\put(300,210){\fbox{${\mathcal A}_{\rm CP}$ in \% }}
	\put(362,122){\scriptsize{-15}}
	\put(320,68){\scriptsize{-12}}
	\put(340,60){\scriptsize{-9}}
	\put(360,55){\scriptsize{-6}}
	\put(380,43){\scriptsize{-3}}
	\put(300,87){\scriptsize{-9}}
	\put(292,96){\scriptsize{-6}}
	\put(283,110){\scriptsize{-3}}
	\put(293,160){\scriptsize{0}}
	\put(245,-8){Fig.~\ref{fig3}b }
\end{picture}
\vspace*{.25cm}
 \caption{Contour lines of $\sigma$ and ${\mathcal A}_{\rm CP}$
in the $|\mu|$-$M_2$ plane for
 $\varphi_{A_{\tau}}=0.5\pi$, $\varphi_{M_1}=\varphi_{\mu}=0$,
$A_{\tau}=1$~TeV, $\tan \beta=5$ and $(P_{e^-},P_{e^+})=(-0.8,0.6)$.
The blank area outside the area of the contour lines is kinematically
forbidden since here either $\sqrt{s}<m_{\CH_1}+m_{\CH_2}$ or
$m_{\T_1}+m_{\tau}>m_{\CH_2}$. The gray area is excluded since
$m_{\CH^{\pm}_1}<104$~GeV.
\label{fig3}}
\end{figure}

Figs.~\ref{fig3}a and \ref{fig3}b show, 
for $\varphi_{A_{\tau}}=0.5\pi$ and $\varphi_{M_1}=\varphi_{\mu}=0$,
the $|\mu|$-$M_2$ dependence of  
the cross section $\sigma$ and the asymmetry 
${\mathcal A}_{\rm CP}$, respectively.
The asymmetry reaches values up to $-15 \%$ due to the 
large value of $|A_{\tau}|=1$ TeV and the choice of 
the beam polarization $(P_{e^-},P_{e^+})=(-0.8,0.6)$. 
This choice also enhances the cross section, which reaches 
values of more than $100$~fb. The gray shaded area excludes 
chargino masses $m_{\CH^{\pm}_1}<104$~GeV.
In the blank area either the sum of the masses of the produced neutralinos
exceeds $\sqrt{s}=500$ GeV or the two-body decay
$\CH^0_2\to \T_1^+\tau^- $ is not open.

\begin{figure}
\begin{picture}(120,220)(0,0)
\put(0,0){\includegraphics{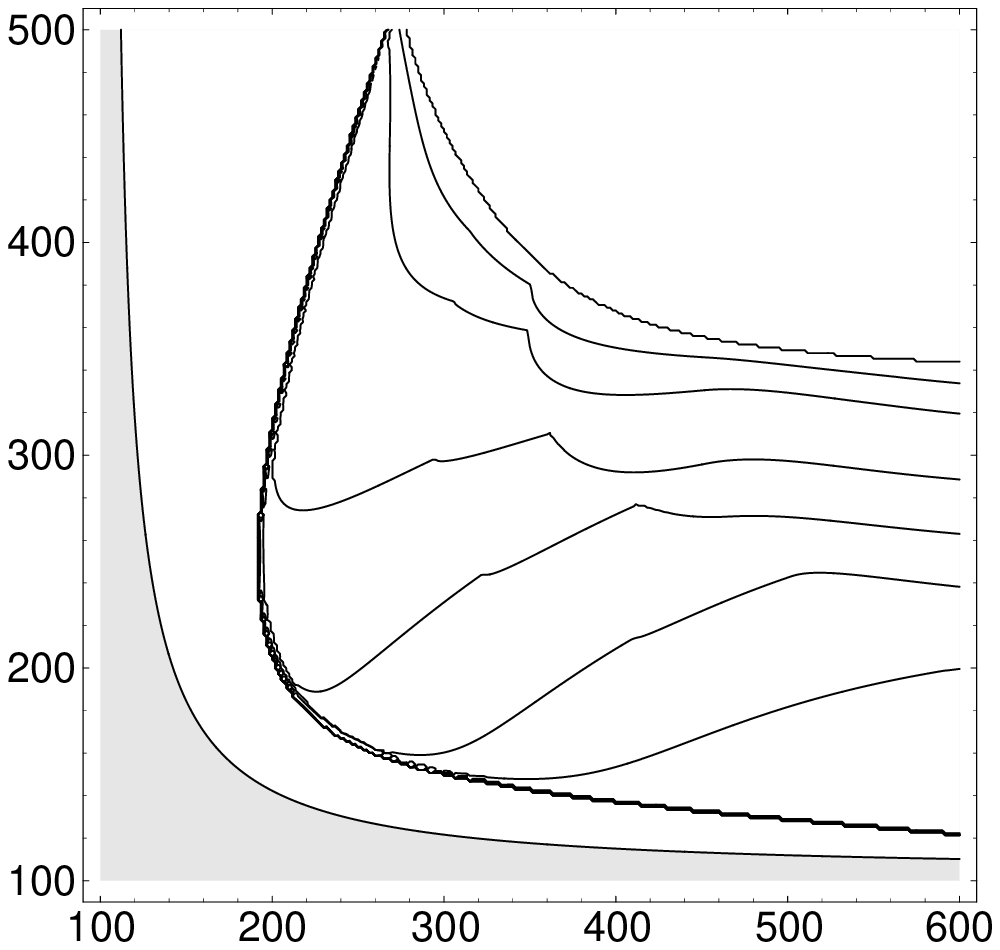}}
   \put(165,-8){$|\mu|$/GeV}
	\put(-5,210){$M_2$/GeV}
	\put(42,210){\fbox{$\sigma(e^+e^-\to \CH^0_1\T_1^+\tau^-)$ in fb }}
		\put(170,49){\footnotesize{150}}
	\put(135,64){\footnotesize{100}}
	\put(102,74){\footnotesize{50}}
	\put(82,94){\footnotesize{25}}
	\put(85,127){\footnotesize{10}}
	\put(86,152){\footnotesize{5}}
	\put(20,-8){Fig.~\ref{fig4}a}
\put(225,0){\includegraphics{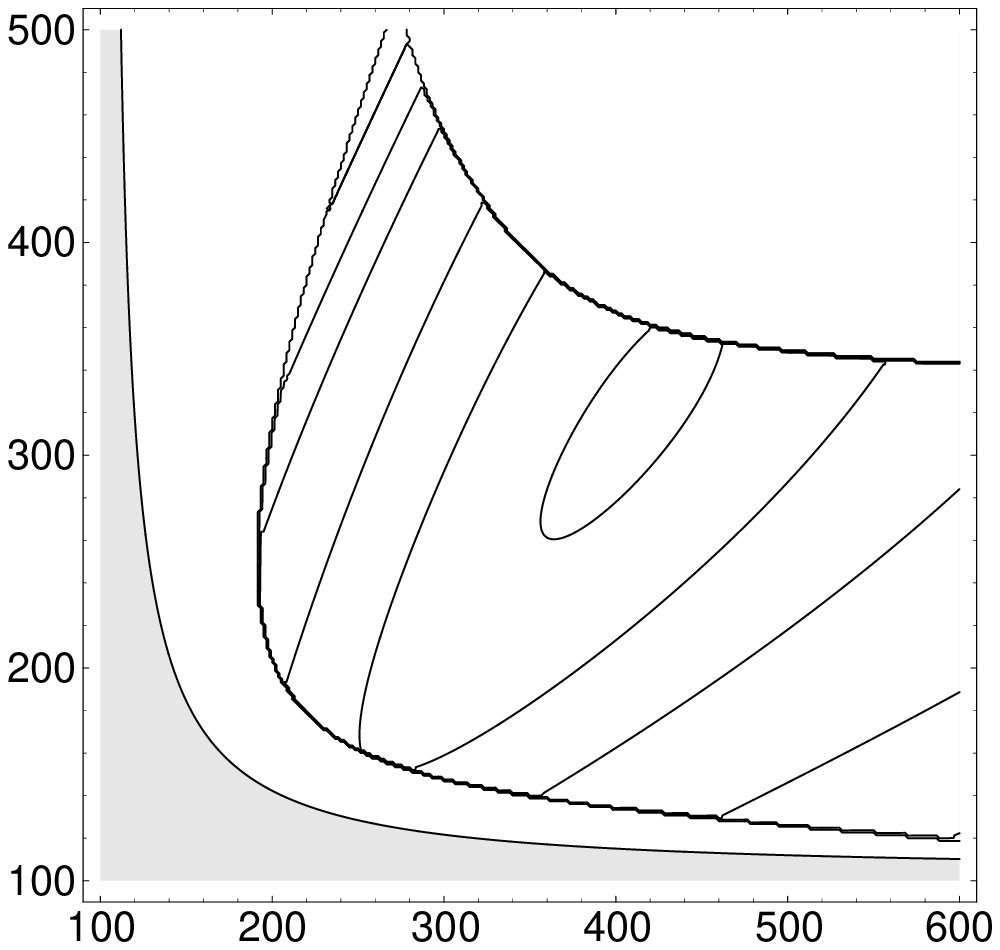}}
   \put(390,-8){$|\mu|$/GeV}
	\put(225,210){$M_2$/GeV}
	\put(300,210){\fbox{${\mathcal A}_{\rm CP}$ in \% }}
	\put(340,105){\footnotesize{-38}}
	\put(360,73){\footnotesize{-30}}
	\put(385, 65){\footnotesize{-20}}
	\put(400, 40){\footnotesize{-10}}
	\put(320,110){\footnotesize{-30}}
	\put(305,125){\footnotesize{-20}}
	\put(280,110){\footnotesize{-10}}
	\put(295,157){-5}
	\put(245,-8){Fig.~\ref{fig4}b }
\end{picture}
\vspace*{.25cm}
 \caption{Contour lines of $\sigma$  and ${\mathcal A}_{\rm CP}$
in the $|\mu|$-$M_2$ plane for
 $\varphi_{M_1}=0.5\pi$, $\varphi_{A_{\tau}}=\varphi_{\mu}=0$,
 $A_{\tau}=250$~GeV, $\tan \beta=5$ and $(P_{e^-},P_{e^+})=(-0.8,0.6)$.
The blank area outside the area of the contour lines is kinematically
forbidden since here either $\sqrt{s}<m_{\CH_1}+m_{\CH_2}$ or
$m_{\T_1}+m_{\tau}>m_{\CH_2}$. The gray area is excluded since
$m_{\CH^{\pm}_1}<104$~GeV.
\label{fig4}}
\end{figure}

For $\varphi_{M_1}= 0.5\pi$ and
$\varphi_{\mu}=\varphi_{A_{\tau}}=0$ we show in Figs.~\ref{fig4}a,b 
the contour lines of $\sigma$  and ${\mathcal A}_{\rm CP}$
in the $|\mu|$-$M_2$ plane, respectively.
It is remarkable that in a large region the asymmetry is larger than
-10\% and reaches values up to -40\% while also the cross section 
is large.
Unpolarized beams would reduce the largest values of 
$\sigma$ by a factor 3, whereas ${\mathcal A}_{\rm CP}$
would only be marginally reduced.

\begin{figure}
\begin{picture}(120,220)(0,0)
\put(0,0){\includegraphics{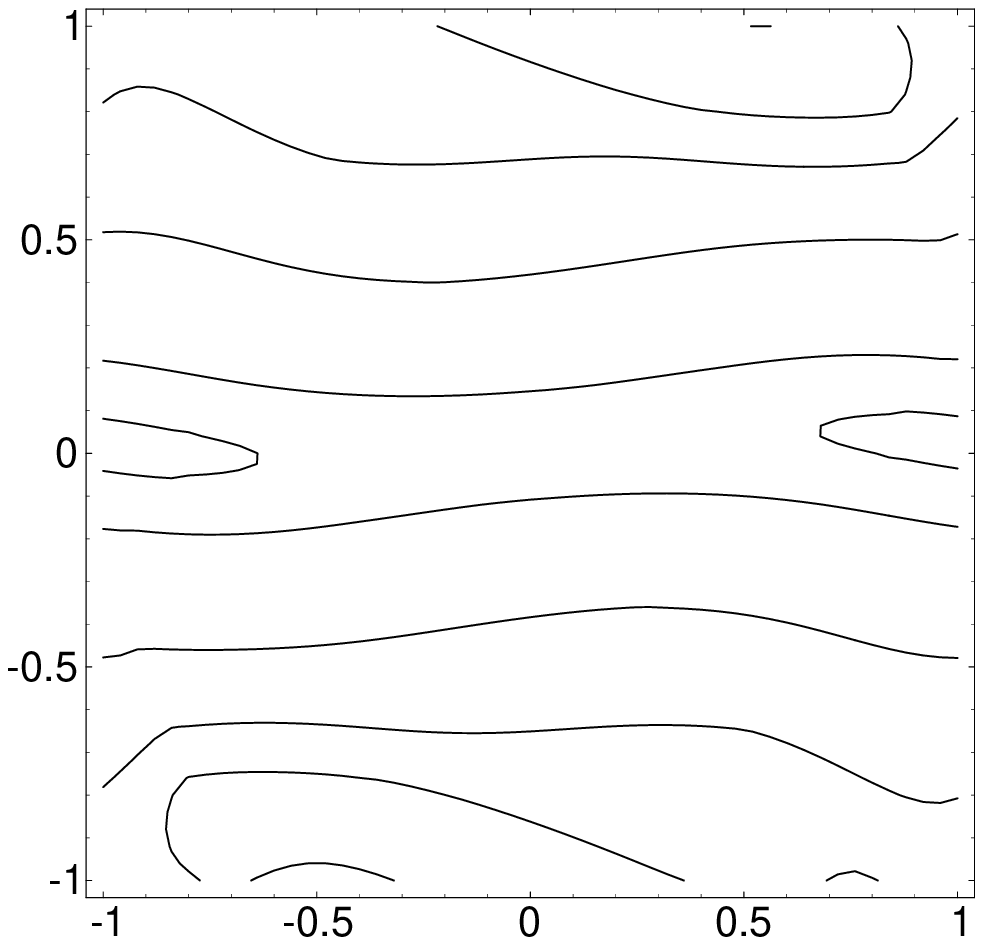}}
   \put(165,-8){$\varphi_{\mu}/\pi$}
	\put(-5,210){$\varphi_{M_1}/\pi$ }
	\put(42,210){\fbox{$\sigma(e^+e^-\to \CH^0_1\T_1^+\tau^-)$ in fb }}
	\put(125,30){\footnotesize{15}}
	\put(130,55){\footnotesize{10}}
	\put(110,74){\footnotesize{5}}
	\put(100,96){\footnotesize{1}}
	\put(100,120){\footnotesize{1}}
	\put(90,143){\footnotesize{5}}
	\put(70,168){\footnotesize{10}}
	\put(140,180){\footnotesize{15}}
	\put(30,105){\footnotesize{0.2}}
	\put(185,105){\footnotesize{0.2}}
		\put(20,-8){Fig.~\ref{fig5}a}
\put(225,0){\includegraphics{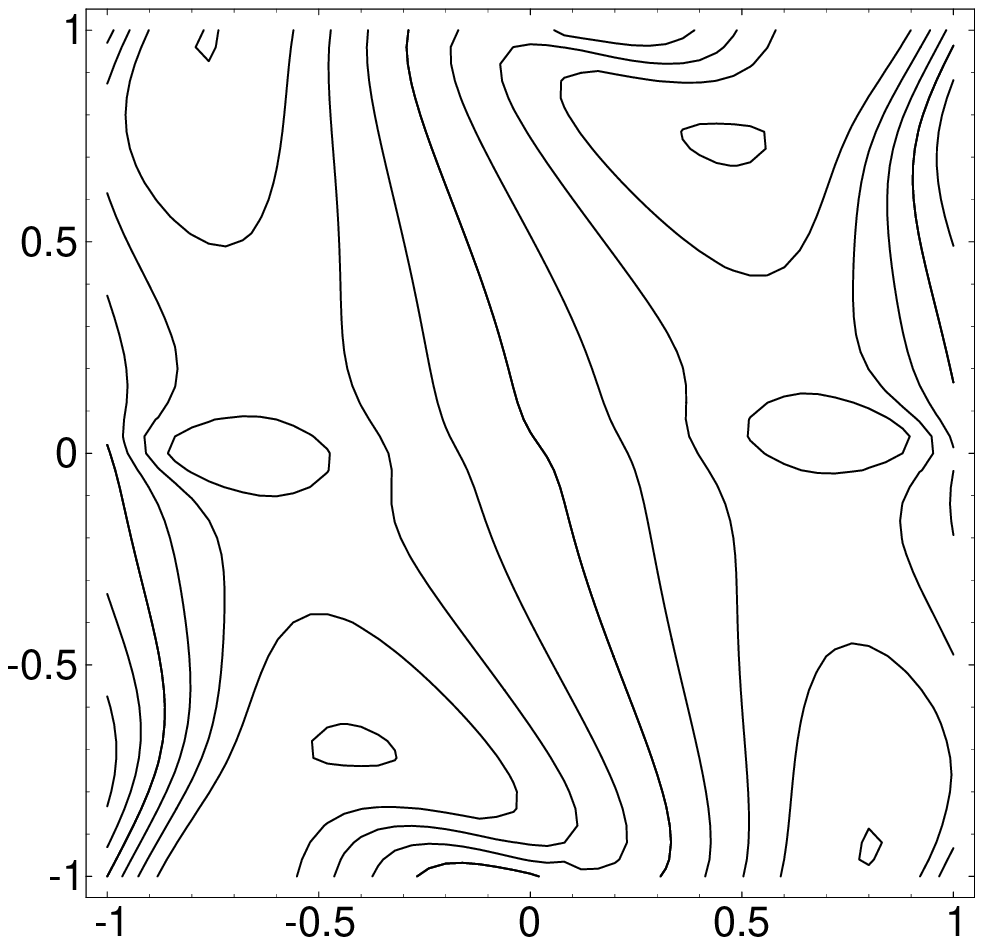}}
   \put(390,-8){$\varphi_{\mu}/\pi$}
	\put(225,210){$\varphi_{M_1}/\pi$ }
	\put(300,210){\fbox{${\mathcal A}_{\rm CP}$ in \% }}
	\put(296,45){\scriptsize{65}}
	\put(290,66){\scriptsize{45}}
	\put(277,80){\scriptsize{30}}
	\put(254,85){\scriptsize{0}}
	\put(275,103){\scriptsize{45}}
	\put(290,130){\scriptsize{30}}
	\put(251,135){\scriptsize{15}}
	\put(265,153){\scriptsize{45}}

	\put(323,80){\scriptsize{15}}
	\put(340,90){\scriptsize{0}}
	\put(350,100){\scriptsize{-15}}
	\put(403,30){\scriptsize{-65}}
	\put(395,58){\scriptsize{-45}}
	\put(375,83){\scriptsize{-30}}
	\put(393,107){\scriptsize{-45}}
	\put(391,125){\scriptsize{-30}}
	\put(405,135){\scriptsize{-15}}
	\put(420,125){\scriptsize{0}}
	\put(373,146){\scriptsize{-45}}
	\put(371,167){\scriptsize{-65}}
	\put(245,-8){Fig.~\ref{fig5}b }
\end{picture}
\vspace*{.25cm}
 \caption{Contour lines of $\sigma$  and ${\mathcal A}_{\rm CP}$
	 in the $\varphi_{\mu}$-$\varphi_{M_1}$ plane for
$M_2=400$ GeV, $|\mu|=300$ GeV, 
$\tan \beta=5$, $\varphi_{A_{\tau}}=0$, $A_{\tau}=250$~GeV 
and $(P_{e^-},P_{e^+})=(-0.8,0.6)$.
\label{fig5}}
\end{figure}

For $|\mu|=300$~GeV and $M_2=400$~GeV, we show in 
Figs.~\ref{fig5}a,b contour lines of 
$\sigma$ and ${\mathcal A}_{\rm CP}$, respectively,  
in the $\varphi_{\mu}$-$\varphi_{M_1}$ plane.
As can be seen the asymmetry ${\mathcal A}_{\rm CP}$ is very sensitive
to variations of the phases $\varphi_{M_1}$ and $\varphi_{\mu}$.
Even for small phases, ${\mathcal A}_{\rm CP}$ can be sizable.
Small values of the phases, especially of $\varphi_{\mu}$,
are suggested by constraints on
electron and neutron electric dipole moments (EDMs) \cite{edmsexp} 
for a typical SUSY scale of the order of a few 100 GeV
(for a review see, e.g., \cite{edmstheo}).


\begin{figure}
			\begin{minipage}{0.47\textwidth}
 \begin{picture}(100,230)(0,0)
	 \put(0,0){\includegraphics{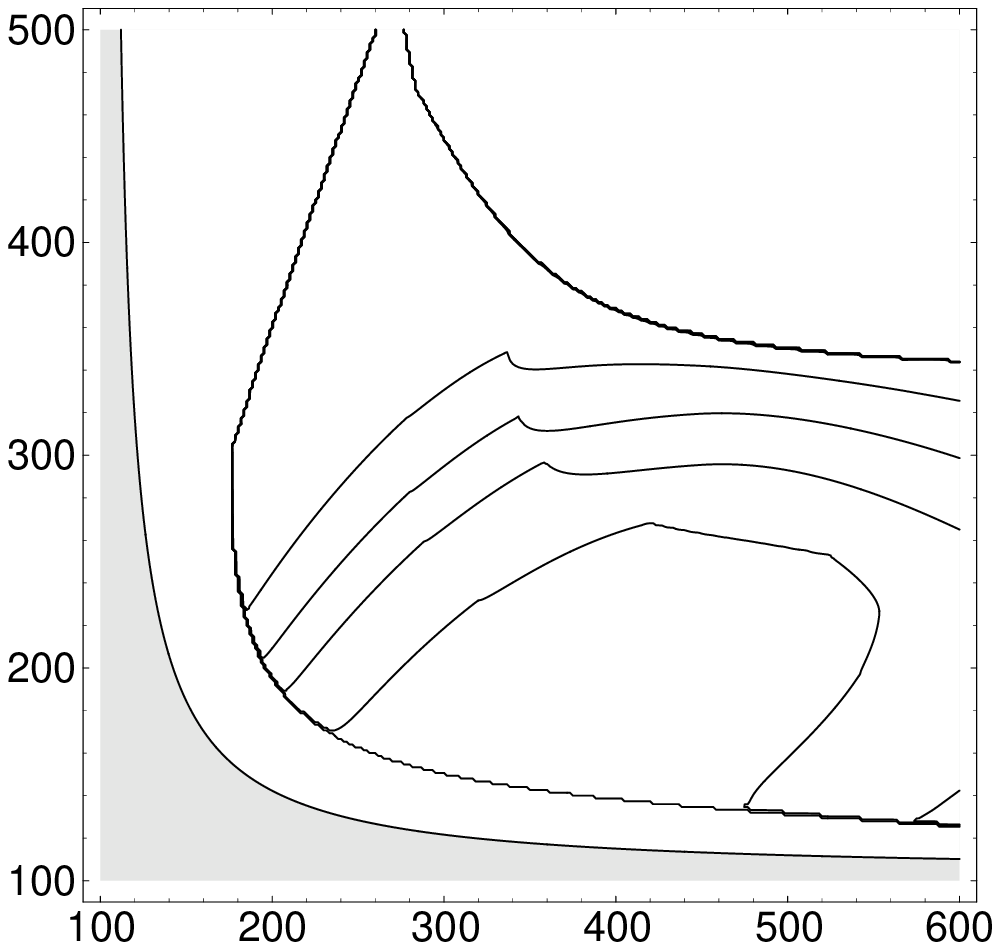}}
	\put(80,210){\fbox{sensitivity $S$}}
	\put(170,-5){$|\mu|$/GeV}
	\put(0,210){$M_2$/GeV}
	\put(121,80){\footnotesize{0.15}}
	\put(137,97){\footnotesize{0.3}}
	\put(119,114){\footnotesize{0.5}}
	\put(100,126){\footnotesize{1}}
 \end{picture}
\vspace*{.25cm}
\caption{Contour lines of  $S$
for $\varphi_{A_{\tau}}=0.2\pi$,
$\varphi_{M_1}=\varphi_{\mu}=0$, 
$A_{\tau}=1$~TeV, $\tan \beta=5$  and 
$(P_{e^-},P_{e^+})=(-0.8,0.6)$.
The blank area outside the area of the contour lines is kinematically
forbidden since here either $\sqrt{s}<m_{\CH_1}+m_{\CH_2}$ or
$m_{\T_1}+m_{\tau}>m_{\CH_2}$. The gray area is excluded since
$m_{\CH^{\pm}_1}<104$~GeV.
\label{fig6}}
\end{minipage}
\hspace*{0.5cm}
\begin{minipage}{0.47\textwidth}
 \begin{picture}(100,230)(0,0)
	\put(0,0){\includegraphics{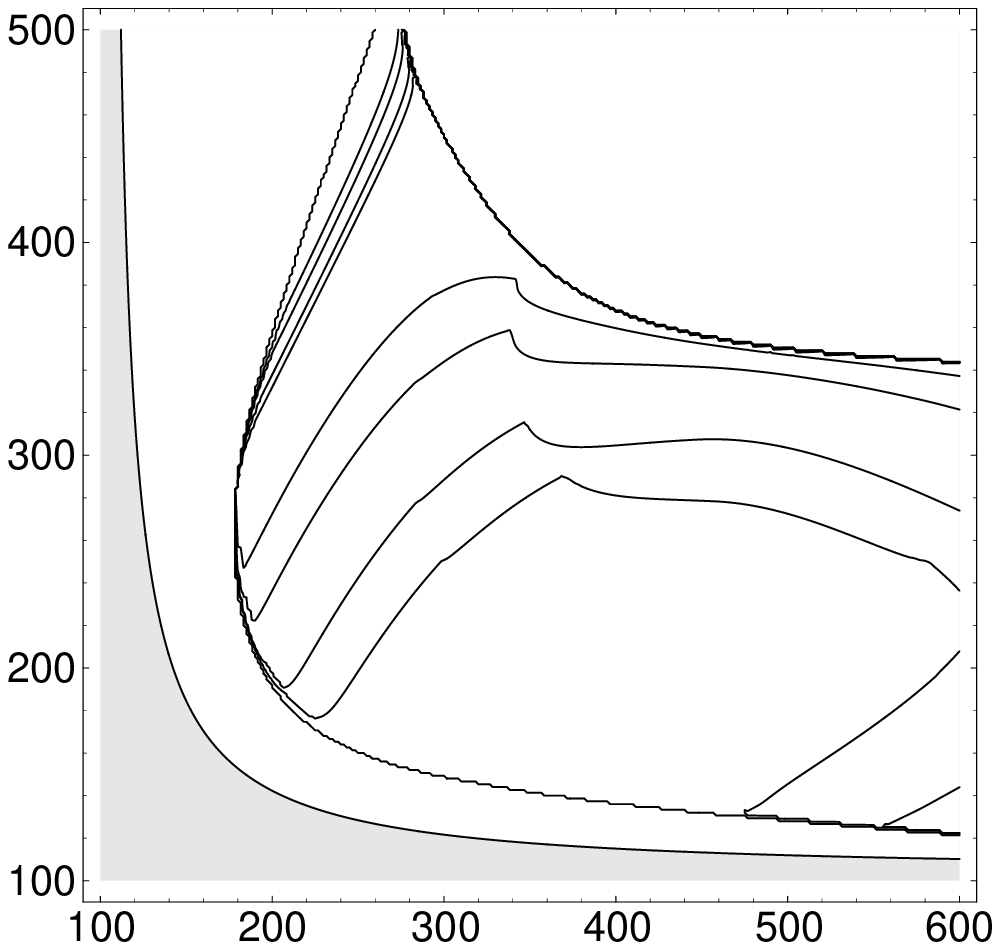}}
	\put(80,210){\fbox{sensitivity $S$}}
	\put(170,-5){$|\mu|$/GeV}
	\put(0,210){$M_2$/GeV}
	\put(95,85){\footnotesize{0.1}}
	\put(125,112){\footnotesize{0.15}}
	\put(90,118){\footnotesize{0.3}}
	\put(88,143){\footnotesize{0.5}}
 \end{picture}
\vspace*{.25cm}
\caption{Contour lines of  $S$
for $\varphi_{M_1}=0.2\pi$,
$\varphi_{A_{\tau}}=\varphi_{\mu}=0$,
$A_{\tau}=250$~GeV, $\tan \beta=5$  and
$(P_{e^-},P_{e^+})=(-0.8,0.6)$.
The blank area outside the area of the contour lines is kinematically
forbidden since here either $\sqrt{s}<m_{\CH_1}+m_{\CH_2}$ or
$m_{\T_1}+m_{\tau}>m_{\CH_2}$. The gray area is excluded since
$m_{\CH^{\pm}_1}<104$~GeV.
\label{fig7}}
\end{minipage}
\vspace*{0.5cm}
\end{figure}

The polarization of the $\tau$ is analyzed through its decay distributions. 
The sensitivities for measuring the polarization of the $\tau$ 
lepton for the various decay modes are quoted in \cite{davier}.
The numbers quoted are for an ideal detector and 
for longitudinal $\tau$ polarization and 
it is expected that the sensitivities for transversely polarized $\tau$ leptons
are somewhat smaller.
Combing informations of all $\tau$ decay modes
a sensitivity of $S=0.35$ \cite{atwood} has been obtained. 
Following \cite{davier}, the relative statistical error of  
$P_2$ (and of $\bar P_2$ analogously)
can be calculated as $\delta P_2= \Delta P_2/|P_2|=
\sigma^s/(S|P_2|\sqrt{N})$, for $\sigma^s$ standard deviations,
where $N=\sigma{\mathcal L}$ is the number of events with
integrated luminosity $\mathcal L$ and cross section
$\sigma = \sigma_p(e^+e^-\to \CH^0_1\CH^0_2)\times 
BR(\CH^0_2\to\T_1^+\tau^-)$.
Then for ${\mathcal A}_{\rm CP}$, Eq.~\rf{eq:asy},
it follows that $\Delta {\mathcal A}_{\rm CP}=\Delta P_2/\sqrt{2}$.
We show in Fig.~\ref{fig6} the contour lines of the sensitivity 
$S=\sqrt{2}/(|{\mathcal A}_{\rm CP}|\sqrt{N})$ which is needed to measure 
${\mathcal A}_{\rm CP}$ at $95\%$ CL ($\sigma^s=2$) for
${\mathcal L}=$500 fb$^{-1}$, for the parameters
$\varphi_{A_{\tau}}=0.2\pi$, $\varphi_{M_1}=\varphi_{\mu}=0$.
In Fig.~\ref{fig7} we show the contour lines of the sensitivity 
$s$ for the parameters
$\varphi_{M_1}=0.2\pi$ and $\varphi_{\mu}=\varphi_{A_{\tau}}=0$.
It is interesting to note that in a large region in the $|\mu|$-$M_2$
plane in Figs.~\ref{fig6} and \ref{fig7} we obtain a sensitivity $S<0.35$, 
which means that the asymmetries can be measured at $95\%$ CL.

\section{Summary and conclusion \label{conclusion}}

We have proposed and analyzed the CP odd asymmetry
${\mathcal A}_{\rm CP}$ in Eq.~\rf{eq:asy} in  neutralino production 
$e^+e^- \to\tilde\chi^0_i \tilde\chi^0_j$
and the subsequent  two-body decay of one  neutralino
$\tilde\chi^0_i \to \tilde\tau_k^{\pm}  \tau^{\mp}$. 
The asymmetry is due to the transverse $\tau^{\mp}$ polarization, 
which is non-vanishing if CP-violating phases of the 
the trilinear scalar coupling parameter  $A_{\tau}$ and/or
the gaugino and higgsino mass parameters $M_1$, $\mu$ are present. 
The asymmetry occurs already at tree level and is due to spin effects in the 
neutralino production and decay process. 
In a numerical study for 
$e^+e^- \to\tilde\chi^0_1 \tilde\chi^0_2$
and neutralino decay $\tilde\chi^0_2 \to \tilde\tau_1^{\pm} \tau^{\mp}$
we have shown that the asymmetry can be as large as 60\%.
It can be sizeable even for small phases of $\mu$ and $M_1$,
which is suggested by the experimental limits on EDMs. 
Depending on the MSSM scenario, the asymmetry should be 
accessible in future electron-positron linear collider experiments 
in the 500 GeV range. Longitudinally polarized electron and positron
beams can considerably enhance both the asymmetry
and the production cross section.

\vskip10mm
\section*{Acknowledgements}

This work is supported by the `Fonds zur
F\"orderung der wissenschaftlichen Forschung' (FWF) of Austria, projects
No. P13139-PHY and No. P16592-N02, by the European Community's 
Human Potential Programme
under contract HPRN--CT--2000--00149 and
by the Deutsche Forschungsgemeinschaft
(DFG) under contract Fr 1064/5-1.

\end{document}